\documentclass[aps,prd,showpacs,twocolumn]{revtex4}
\usepackage{graphicx}
\usepackage{amsmath}
\usepackage{bm}
\usepackage{color}
\usepackage{amssymb}

\def\be{\begin{equation}}
\def\ee{\end{equation}}
\def\bea{\begin{eqnarray}}
\def\eea{\end{eqnarray}}

\def\lp{\left(}
\def\rp{\right)}

\def\R{\mathcal{R}}

\begin{document}

%-----------------------------------------------------------------------
\title{Galactic rotation curves in  hybrid metric-Palatini gravity}
%-----------------------------------------------------------------------

\author{Salvatore Capozziello$^{1.2}$}\email{capozzie@na.infn.it}
\author{Tiberiu Harko$^3$}\email{t.harko@ucl.ac.uk}
\author{Tomi S. Koivisto$^{4}$}\email{tomi.koivisto@fys.uio.no}
\author{Francisco S.N.~Lobo$^{5}$}\email{flobo@cii.fc.ul.pt}
\author{Gonzalo J. Olmo$^{6}$}\email{gonzalo.olmo@csic.es}

\affiliation{$^1$Dipartimento di Fisica, Universit\`{a} di Napoli "Federico II", Napoli, Italy
$^2$INFN Sez. di Napoli, Compl. Univ. di Monte S. Angelo, Edificio G, Via Cinthia, I-80126, Napoli, Italy}
\affiliation{$^3$Department of Mathematics, University College London, Gower Street, London WC1E 6BT, United Kingdom}
\affiliation{$^{4}$  Institute of Theoretical Astrophysics, University of
  Oslo, P.O.\ Box 1029 Blindern, N-0315 Oslo, Norway}
\affiliation{$^5$Centro de Astronomia e Astrof\'{\i}sica da Universidade de Lisboa, Campo Grande, Ed. C8 1749-016 Lisboa, Portugal}
\affiliation{$^6$Departamento de F\'{i}sica Te\'{o}rica and IFIC, Centro Mixto Universidad de
Valencia - CSIC. Universidad de Valencia, Burjassot-46100, Valencia, Spain}

%-----------------------------------------------------------------------
\begin{abstract}

Generally, the dynamics of test particles around galaxies, as well as the corresponding mass deficit, is explained by postulating the existence of a hypothetical dark matter. In fact, the behavior of the rotation curves shows the existence of a constant velocity region, near the baryonic matter distribution, followed by a quick decay at large distances. In this work, we consider the possibility that the behavior of the rotational velocities of test particles gravitating around galaxies can be explained within the framework of the recently proposed hybrid metric-Palatini gravitational theory. The latter is constructed by modifying the metric Einstein-Hilbert action with an $f(R)$ term in the Palatini formalism. It was shown that the theory unifies local constraints and the late-time cosmic acceleration, even if the scalar field is very light. In the intermediate galactic scale, we show explicitly that in the hybrid metric-Palatini model the tangential velocity can be explicitly obtained as a function of the scalar field of the equivalent scalar-tensor description. The model predictions are compared model with a small sample
of rotation curves of low surface brightness galaxies, respectively, and a good
agreement between the theoretical rotation curves and the observational data is found. The possibility of constraining the form of the scalar field and the parameters of the model by using the stellar velocity dispersions is also analyzed. Furthermore, the Doppler velocity shifts are also obtained in terms of the scalar field. All the physical and geometrical quantities and the numerical parameters in the hybrid metric-Palatini model can be expressed in terms of observable/measurable parameters, such as the tangential velocity, the baryonic mass of the galaxy, the Doppler frequency shifts, and the stellar dispersion velocity, respectively. Therefore, the obtained results open the possibility of testing the hybrid metric-Palatini gravitational models  at the galactic or extra-galactic scale by using direct astronomical and astrophysical observations.

\end{abstract}
%-----------------------------------------------------------------------


\date{\today}

\maketitle

%-----------------------------------------------------------------------
\section{Introduction}
%-----------------------------------------------------------------------

A long outstanding challenge in modern astrophysics is the intriguing observational behavior of the galactic rotation curves, and the mass discrepancy in clusters of galaxies. Both these observations suggest the existence  of a (non or weakly interacting) form of dark matter at galactic and extra-galactic scales. Indeed, according to Newton's theory of gravitation, at the boundary of the luminous matter, the rotation curves of test particles gravitating around galaxies or galaxy clusters should show a Keplerian decrease of the tangential rotational speed $v_{tg}$ with the distance $r$, so that $v_{tg}^2\propto M(r)/r$, where $M(r)$ is the dynamical mass within the radius $r$. However, the observational evidence indicates rather flat rotation curves \cite{dm, BT08}. The tangential rotational velocities $v_{tg}$ increase near the galactic center, as expected, but then intriguingly remain approximately constant at a value of $v_{tg\infty }\sim 200-300$ km/s. Therefore, observations provide a general mass profile of the form $M(r)\approx rv_{tg\infty }^2/G$ \cite{dm, BT08}. Consequently, the mass around a galaxy increases linearly with the distance from the center, even at large distances, where very little or no luminous matter can be observed.

As mentioned above, the observed behavior of the galactic rotation curves is explained by assuming the existence of some dark (invisible) matter, distributed in a spherical halo around the galaxies, and interacting only gravitationally with ordinary matter. The dark matter is usually described as a pressureless and cold medium. A recently proposed model considered the possibility that dark matter is a mixture of two non-interacting perfect fluids, with different four-velocities and thermodynamic parameters. The two-fluid model can be described as an effective single anisotropic fluid, with distinct radial and tangential pressures \cite{Harko:2011nu}. In fact, many possible candidates for non-luminous dark matter have been proposed in the literature ranging from $10^6$ solar mass black holes, running through low mass stars to $10^{-6}$ eV axions, although the most popular being the weakly interacting massive particles (WIMPs) (for a review of the particle physics aspects of dark matter see \cite{OvWe04}). Indeed, the interaction cross section of WIMPs with normal baryonic matter, although practically negligible, is expected to be non-zero, and therefore there is a possibility of detecting them directly. Nevertheless, despite several decades of intense experimental and observational effort, there is presently still no direct evidence of dark matter particles \cite{Wexp}.

However, it is important to emphasize that the masses in galaxies and clusters of galaxies are deduced from the observed distances and velocities of the system under consideration. These relationships are based on Newton's laws of dynamics, thus the Newtonian dynamical masses of galactic systems are not consistent with the observed masses. Indeed, Newton's laws have proven extremely reliable in describing local phenomena, that there is an overwhelming tendency to apply them in the intermediate galactic scales. Note that the mass discrepancy is interpreted as evidence for the existence of a {\it missing mass}, i.e., dark matter in galactic systems. Therefore, one cannot {\it a priori} exclude the possibility that Einstein's (and Newtonian) gravity breaks down at the galactic or extra-galactic scales. Indeed, a very promising way to explain the recent observational data \cite{Ri98,PeRa03} on the recent acceleration of the Universe and on dark matter is to assume that at large scales Einstein's general relativity, as well as its Newtonian limit,  breaks down, and a more general action describes the gravitational field. Several theoretical models, based on a modification of Newton's law or of general relativity,  including Modified Newtonian Orbital Dynamics (MOND) \cite{Mond}, scalar fields or long range coherent fields coupled to gravity \cite{scal1}, brane world models \cite{brane}, Bose-Einstein condensates \cite{bose},  modified gravity with geometry-matter coupling \cite{gmc}, non-symmetric theories of gravity \cite{scal3}, and Eddington-inspired Born-Infeld gravity \cite{Ed} have been used to model galactic ``dark matter''.

From a theoretical point of view a very attractive possibility is to modify the form of the Einstein-Hilbert Lagrangian, so that such a modification could naturally explain dark matter and dark energy, without  resorting to any exotic forms of matter. The simplest extension of the Einstein-Hilbert action consists in modifying the geometric part of the action, through the substitution of the Ricci scalar with a generic function $f(R)$. This change in the action introduces  higher order terms in the gravitational field action. The
so-called $f(R)$ gravitational theories were first proposed in \cite{carroll}, and later used  to
find a non-singular isotropic de Sitter type cosmological solution  \cite{staro}. Detailed reviews of $f(R)$ theories can be found, for instance, in \cite{rev}. The most
serious difficulty of  $f(R)$ theories is that, in general, these theories cannot pass the
standard Solar System tests \cite{badfR}. However, there exists some classes of theories that can solve
this problem \cite{goodfR}. Using phase space analysis of the specific
involved gravitational model, it was shown that $f(R)$ theories, in general, can explain the evolution of the
Universe, from a  matter dominated early epoch up to the present, late-time self accelerating phase \cite{phasefR}.
 $f(R)$ type gravity theories can be generalized by including the function $f(R)$ in the bulk action of the brane-world theories \cite{shahab}.  It has been shown in \cite{dynam} that this type of generalized brane world theories can describe a Universe beginning with a matter-dominated era, and ending in an accelerated expanding phase. The classical tests of this theory were  considered in \cite{zahra}.

 On the other hand, $f(R)$ theories have also been studied in the Palatini approach, where the metric and the connection are regarded as independent fields \cite{P}. In fact, these approaches are certainly equivalent in the context of general relativity, i.e., in the case of the linear Einstein-Hilbert action. On the other hand, for  a general $f(R)$ term in the action, they seem to provide completely different theories, with very distinct field equations. The Palatini variational approach, for instance, leads to second order differential field equations, while the resulting field equations in the metric approach are fourth order coupled differential equations. These differences also extend to the observational aspects. All the Palatini $f(R)$ models aimed at explaining the cosmic speedup studied so far  lead to microscopic matter instabilities, and to unacceptable features in the evolution patterns of cosmological perturbations \cite{P1}.

Hence, in order to cure some of the pathologies of the $f(R)$ gravity models in both their metric and Palatini formulation,  a novel approach was recently proposed \cite{fX}, that consists of adding to the Einstein-Hilbert Lagrangian an $f(R)$ term constructed within the framework of the  Palatini formalism. Using the respective dynamically equivalent scalar-tensor representation, even if the scalar field is very light, the theory can pass the Solar System observational constraints. Therefore the long-range scalar field is able to modify the cosmological and galactic dynamics, but leaves the Solar System unaffected. The absence of instabilities in perturbations was also verified, and explicit models, which are consistent with local tests and lead to the late-time cosmic acceleration, were also found. Furthermore, the cosmological applications of the hybrid metric-Palatini gravitational theory were investigated in \cite{fX1}, where specific criteria to obtain the cosmic acceleration were analyzed, and the field equations were formulated as a dynamical system. Indeed, several classes of dynamical cosmological solutions, depending on the functional form of the effective scalar field potential, describing both accelerating and decelerating Universes, were explicitly obtained. The cosmological perturbation equations were also derived and applied to uncover the nature of the propagating scalar degree of freedom and the signatures these models predict in the large-scale structure. The general conditions for wormhole solutions according to the null energy condition violation at the throat in the hybrid metric-Palatini gravitational theory were also presented in \cite{fX2}.

A new approach to modified gravity which generalizes the  hybrid metric-Palatini gravity was introduced in \cite{Bo}. The gravitational action is taken to depend on a general function of both the metric and Palatini curvature scalars. The dynamical equivalence with a non-minimally coupled bi-scalar field gravitational theory was proved. The evolution of the cosmological solutions in this model was studied by using dynamical systems techniques.  In \cite{TT} a method was developed to analyse the field content of "hybrid" gravity theories whose actions involve both the independent Palatini connection and the metric Levi-Civita connection, and, in particular, to determine whether the propagating degrees of freedom are ghosts or tachyons. New types of second, fourth and sixth order derivative gravity theories were investigated, and from this analysis it follows that the metric-Palatini theory is singled out as a viable class of ``hybrid'' extensions of General Relativity. In addition to this, the stability of the Einstein static Universe was analysed in \cite{Boehmer:2013oxa}. In the latter, by considering linear homogeneous perturbations, the stability regions of the Einstein static universe were parameterized by the first and second derivatives of the scalar potential, and it was explicitly shown that a large class of stable solutions exists in the respective parameter space. For a brief review of the hybrid metric-Palatini theory, we refer the reader to \cite{IJMPD}.

Thus, the hybrid metric-Palatini theory opens up new possibilities to approach, in the same theoretical framework, the problems of both dark energy and dark matter. In \cite{Capozziello:2012qt}, the generalized virial theorem in the scalar-tensor representation of the hybrid metric-Palatini gravity was analysed. More specifically, taking into account the relativistic collisionless Boltzmann equation, it was shown that the supplementary geometric terms in the gravitational field equations provide an effective contribution to the gravitational potential energy. Indeed, it was shown that the total virial mass is proportional to the effective mass associated with the new terms generated by the effective scalar field, and the baryonic mass. This shows that the geometric origin in the generalized virial theorem may account for the well-known virial theorem mass discrepancy in clusters of galaxies. In addition to this, the astrophysical applications of the model were considered and it was shown that the model predicts that the mass associated to the scalar field and its effects extend beyond the virial radius of the clusters of galaxies. In the context of the galaxy cluster velocity dispersion profiles predicted by the hybrid metric-Palatini model, the generalized virial theorem can be an efficient tool in observationally testing the viability of this class of generalized gravity models. Thus, hybrid metric-Palatini gravity provides an effective alternative to the dark matter paradigm of present day cosmology and astrophysics.

In this latter context, it is the purpose of the present paper to investigate the possibility that the observed properties of the galactic rotation curves could be explained in the framework of hybrid metric-Palatini gravity, without postulating the existence of the hypothetical dark matter. As a first step in this study, we obtain the expression of the tangential velocity of test particles in stable circular orbits around galaxies. Since we assume that the test particles move on the geodesic lines of the space-time, their tangential velocity is determined only by the radial distance to the galactic center, and the metric, through the derivative of the $g_{00}$ metric component with respect to the radial coordinate. The metric in the outer regions of the galaxy is largely shaped by the energy contained in the effective scalar field of the hybrid metric-Palatini gravitational theory.  Therefore the behavior of the neutral hydrogen gas clouds outside the galaxies, and their flat rotation curves, can be explained by the presence of the scalar field generated in the model.  By using the weak field limit of the gravitational field equations in the hybrid metric-Palatini model we obtain the explicit form of the tangential velocity, and show that the existence of a constant velocity region is possible for some specific values of the model parameters.

Since the observations on the galactic rotation curves are obtained from the Doppler frequency shifts,
we generalize the expression of the frequency shifts by including the effect of the scalar field.  We also
consider the velocity dispersion of the stars in the galaxy, and obtain the stellar velocity dispersion as a
function of the scalar field. Thus, at least in principle, all the basic parameters of the model can be
obtained directly from astronomical observations. The knowledge of the tangential velocity allows the
complete determination of the functional form of the scalar field from the observational data. The basic
parameters of the model can be immediately obtained in the flat rotation curves region, which is
determined by the derivative of the scalar field potential. Hence the functional form of the scalar field
can be obtained exactly, within the weak field limit of the model, for the entire galactic space-time, and
tested at the galactic scale. Therefore, all the physical parameters of the hybrid metric-Palatini
gravitational theory can be either obtained directly, or severely constrained by astronomical
observations.

The present paper is organized as follows. In Section \ref{sec:b}, the field equations of the hybrid
metric-Palatini gravity model, as well as the general fluid representation of the stress-energy tensor are
presented. In Section \ref{sect3}, the tangential velocity of test particles in stable circular orbits are
derived and in Section \ref{sect4}, the tangential velocities of test particles in the galactic halos in the
hybrid model are discussed. The comparison of the theoretical predictions for the rotational velocity and the observational data for four Low Surface Brightness galaxies is considered in Section \ref{sect5}. The velocity dispersion of the stars in the galaxy, representing important observational tests
of the model, as well as the red and blue shifts of the electromagnetic radiation emitted by the gas
clouds are also investigated.  We discuss and conclude our results in Section
\ref{sect6}. In this paper, we use the Landau-Lifshitz \cite{LaLi} sign conventions and definitions of the
geometric quantities.

%-----------------------------------------------------------------------
\section{Field equations of the scalar-tensor version of hybrid metric-Palatini $f(X)$-gravity}
\label{sec:b}
%-----------------------------------------------------------------------

In this Section, we briefly present the basic formalism, for self-completeness and self-consistency, and the field equations of the hybrid metric-Palatini gravitational theory within the equivalent scalar-tensor representation (we refer the reader to \cite{fX,fX1} for more details), and furthermore obtain the perfect fluid form of the stress-energy tensor of the scalar field.

%-----------------------------------------------------------------------
\subsection{Scalar-tensor representation for hybrid metric-Palatini gravity}
%-----------------------------------------------------------------------

The action for hybrid metric-Palatini gravity is obtained by adding an $f(\R)$ term, constructed within the framework  of the Palatini formalism, to the metric Einstein-Hilbert Lagrangian \cite{fX}, and is given by
\begin{equation} \label{action}
S= \frac{1}{2\kappa ^2}\int d^4 x \sqrt{-g} \left[ R + f(\R)\right] +S_m \ ,
\end{equation}
where $\kappa =8\pi G$ is the gravitational coupling constant, and the scalar curvature $\R$, depending on both the metric and an independent
dynamical connection $\hat{\Gamma}^\alpha_{\mu\nu}$, is defined as
\begin{equation}\label{Palatini_curvature}
\R
\equiv  g^{\mu\nu}\R_{\mu\nu} \equiv g^{\mu\nu}\lp
\hat{\Gamma}^\alpha_{\mu\nu , \alpha}
       - \hat{\Gamma}^\alpha_{\mu\alpha , \nu} +
\hat{\Gamma}^\alpha_{\alpha\lambda}\hat{\Gamma}^\lambda_{\mu\nu} -
\hat{\Gamma}^\alpha_{\mu\lambda}\hat{\Gamma}^\lambda_{\alpha\nu}\rp \,.
\end{equation}
$\R_{\mu\nu}$ is the Ricci tensor obtained from the connection $\hat{\Gamma}^\alpha_{\mu\nu}$.

The hybrid metric-Palatini theory may be expressed in a purely scalar-tensor representation, by the following action
\begin{equation} \label{scalar2}
S= \frac{1}{2\kappa^2}\int d^4 x \sqrt{-g} \left[ (1+\phi)R +\frac{3}{2\phi}\nabla_\mu \phi \nabla ^\mu \phi
-V(\phi)\right] +S_m \,,
\end{equation}
which differs fundamentally from the $w=-3/2$ Brans-Dicke theory in the coupling of the scalar to the scalar curvature, where $w$ is the dimensionless Brans-Dicke coupling constant. The variation of this action with respect to the metric tensor gives the field equations
\be\label{einstein_phi}
G_{\mu\nu}=\kappa^2\left(\frac{1}{1+\phi}T_{\mu\nu} +T_{\mu \nu}^{(\phi)}\right),
\ee
where $T_{\mu \nu}$ is the ordinary matter stress-energy tensor, and
\bea
T_{\mu \nu}^{(\phi)}&=&\frac{1}{\kappa ^2}\frac{1}{1+\phi }\Bigg[\nabla_\mu\nabla_\nu\phi -\frac{3}{2\phi}\nabla_\mu\phi\nabla_\nu\phi +\nonumber\\
&&\Bigg(\frac{3}{4\phi}\nabla_\lambda\phi\nabla^\lambda\phi - \Box\phi- \frac{1}{2}V\Bigg)g_{\mu \nu}\Bigg],
\eea
is the stress-energy tensor of the  scalar field.
The variation of the action with respect to the scalar field gives
\be\label{variation_phi}
R - \frac{3}{\phi}\Box\phi +\frac{3}{2\phi^2}\nabla _\mu \phi \nabla ^\mu \phi - \frac{dV}{d\phi} =0 \,.
\ee

Moreover, one can show that the identity
\begin{equation}\label{eq:phi(X)}
2V-\phi \frac{dV}{d\phi}=\kappa^2T+R \,,
\end{equation}
also holds, and that the scalar field $\phi$ is
governed by the second-order evolution equation
\begin{equation}\label{eq:evol-phi}
-\Box\phi+\frac{1}{2\phi}\nabla _\mu \phi \nabla ^\mu
\phi+\frac{1}{3}\phi\left[2V-(1+\phi)\frac{dV}{d\phi}\right]=\frac{\phi\kappa^2}{3}T\,,
\end{equation}
with $T=T_{\mu }^{\mu }$, which is an effective Klein-Gordon equation.

%-----------------------------------------------------------------------
\subsection{The perfect fluid representation of the scalar field stress-energy tensor}
%-----------------------------------------------------------------------

The stress-energy tensor of a fluid can be generally represented as
\be
T_{\mu \nu}=\left(\rho +p\right)u_{\mu }u_{\nu}-pg_{\mu \nu }+q_{\mu }u_{\nu }+q_{\nu }u_{\mu }+S_{\mu \nu},
\ee
where $u_{\mu }$ is the four-velocity of the fluid, $\rho $ and $p$ are the energy density and isotropic pressure, respectively, $q_{\mu }$ is the heat flux, and $S_{\mu \nu}$ is the tensor of the anisotropic dissipative stresses. The heat flux four-vector and the anisotropic stress tensor satisfy the conditions $u^{\mu }q_{\mu }=0$, $S_{\mu }^{\mu }=0$, and $S_{\mu }^{\nu }u_{\nu }=0$, respectively. The four-velocity is normalized so that $u_{\mu }u^{\mu }=1$, and $u_{\nu }\nabla _{\mu }u^{\nu }=0$. By introducing the projection tensor $h^{\mu \nu}=g^{\mu \nu}-u^{\mu \nu}$, with the properties  $g_{\mu \nu}h^{\mu \nu}=3$, $h^{\mu \nu}u_{\nu }=0$, and $h_{\mu \nu}u^{\mu }u^{\nu }=0$, the thermodynamic parameters of the fluid can be obtained from the stress-energy tensor as
\bea\label{ex}
&&\rho =u^{\mu }u^{\nu }T_{\mu \nu }, \qquad  p=-\frac{1}{3}h^{\mu \nu}T_{\mu \nu},
    \nonumber\\
&&q_{\mu }=u^{\alpha }h^{\beta }_{\mu}T_{\alpha \beta }, \qquad
S_{\mu \nu}=h^{\alpha }_{\mu }h^{\beta }_{\nu}T_{\alpha \beta }+ph_{\mu \nu}.
\eea

In order to obtain the perfect fluid representation of the stress-energy tensor of the scalar field in hybrid metric-Palatini gravity, we introduce first the four-velocity of the scalar field as
\be
u^{\mu }_{(\phi)} =\frac{\nabla ^{\mu }\phi }{\sqrt{\nabla _{\alpha }\phi \nabla ^{\alpha }\phi }},
\ee
which satisfies the relation $u^{\mu }_{(\phi)}u_{(\phi)\mu}=1$. Therefore, with the use of Eqs.~(\ref{ex}) we obtain the effective energy density $\rho _{\phi }$ and pressure $p_{\phi }$ in the scalar field description of hybrid metric-Palatini gravity as
\bea
\rho _{\phi} &=&\frac{1}{\kappa ^2}\frac{1}{1+\phi }\Bigg\{\frac{1}{2}\nabla ^{\mu }\phi \nabla _{\mu }\ln \left(\nabla _{\alpha }\phi \nabla ^{\alpha }\phi \right)-\frac{5}{4\phi}\nabla _{\alpha }\phi \nabla ^{\alpha }\phi
\nonumber\\
&&\hspace{-0.5cm}-\frac{1}{3}\left[2V(\phi)-(1+\phi)V'(\phi)\right]+\frac{\kappa ^2}{3}\phi T-\frac{1}{2}V(\phi)\Bigg\},
\eea
\bea
p_{\phi}&=&\frac{1}{\kappa ^2}\frac{1}{1+\phi }\Bigg\{-\sqrt{\nabla _{\alpha }\phi \nabla ^{\alpha }\phi}\;\frac{\Theta }{3}-\frac{1}{4\phi }\nabla _{\alpha }\phi \nabla ^{\alpha }\phi
\nonumber\\
&&\hspace{-0.8cm}+\frac{1}{3}\left[2V(\phi)-(1+\phi)V'(\phi)\right]-\frac{\kappa ^2}{3}\phi T+\frac{1}{2}V(\phi)\Bigg\},
\eea
\be
q^{(\phi )\mu }=\frac{1}{\kappa ^2}\frac{1}{1+\phi }\sqrt{\nabla _{\alpha }\phi \nabla ^{\alpha }\phi}u^{\alpha }_{(\phi) }\nabla _{\alpha }u^{\mu }_{(\phi) },
\ee
\bea
S^{(\phi )\mu \nu}&=&\frac{1}{\kappa ^2}\frac{1}{1+\phi }\times
\nonumber\\
&&\hspace{-0.7cm} \times  \Bigg[\sqrt{\nabla _{\alpha }\phi \nabla ^{\alpha }\phi }\left(\nabla ^{\mu }u^{\nu }_{(\phi )} -u_{(\phi) }^{\alpha}u^{\mu }_{\phi }\nabla _{\alpha }u^{\nu }_{(\phi ) }\right)
\nonumber\\
&& \hspace{-0.7cm} +\left(-\sqrt{\nabla _{\alpha }\phi \nabla ^{\alpha }\phi }\;\frac{\Theta }{3}+\frac{1}{2\phi }\nabla _{\alpha }\phi \nabla ^{\alpha }\phi\right) h^{\mu \nu}\Bigg],
\eea
where $\Theta =\nabla _{\alpha }u^{\alpha }_{(\phi )}$ is the expansion of the fluid. It is interesting to note that the fluid-equivalent stress-energy tensor in hybrid metric-Palatini gravity is not of a perfect fluid form, but contains ``heat transfer'' terms, as well as an anisotropic  dissipative component.

Therefore, the stress-energy of the scalar field can be written in a form equivalent to a general fluid as
\bea
T^{(\phi)\mu \nu}&=&\left(\rho _{\phi }+p_{\phi}\right)u^{\mu }_{(\phi )}u^{\nu }_{(\phi)}-p_{\phi }g^{\mu \nu}+\nonumber\\
&&q^{(\phi )\mu }u^{\nu }_{(\phi)}+q^{(\phi)\nu }u^{\mu }_{(\phi ) }+S^{(\phi )\mu \nu}\,,
\eea
which will be useful in determining the galactic geometry in the context of the tangential velocity curves analysis outlined below.

%-----------------------------------------------------------------------
\section{Stable circular orbits of test particles around galaxies}\label{sect3}
%-----------------------------------------------------------------------

The most direct method for studying the gravitational field inside a spiral galaxy is provided by the galactic rotation
curves. They are obtained by measuring the frequency shifts $z$ of the 21-cm radiation emission from the neutral
hydrogen gas clouds. The 21-cm radiation also originates from stars. The 21-cm background from the epoch of reionization is a promising cosmological probe: line-of-sight velocity fluctuations distort redshift, so brightness fluctuations in Fourier space depend upon angle, which linear theory shows can separate cosmological from astrophysical information (for a recent review see \cite{21cm}).  Instead of using $z$ the resulting redshift is presented by astronomers in terms of a velocity
field $v_{tg}$ \cite{dm}.

In the following, we will assume that the gas clouds behave like test particles, moving in the static and
spherically symmetric geometry around the galaxy. Without a significant loss of generality, we assume that the gas
clouds move in the galactic plane $\theta =\pi /2$, so that their four-velocity is given by $u^{\mu }=\left( \dot{t},
\dot{r},0,\dot{\phi}\right)$, where the overdot stands for derivation with respect to the affine parameter $s$.

The static spherically symmetric metric outside the galactic baryonic mass distribution is given by the following line
element
\begin{equation}
ds^{2}=e^{\nu \left( r\right) }c^2dt^{2}-e^{\lambda \left( r\right)
}dr^{2}-r^{2}\left( d\theta ^{2}+\sin ^{2}\theta d\phi ^{2}\right) ,
\label{line}
\end{equation}
where the metric coefficients $\nu (r)$ and $\lambda (r)$ are functions of the radial coordinate $r$ only. The motion of a
test particle in the gravitational field with the metric given by Eq.~(\ref{line}), is described by the Lagrangian \cite{Nuc01}
\begin{equation}
L=\left[ e^{\nu \left( r\right) }\left( \frac{cdt}{ds}\right)
^{2}-e^{\lambda \left( r\right) }\left( \frac{dr}{ds}\right)
^{2}-r^{2}\left( \frac{d\Omega }{ds}\right) ^{2}\right] ,
\end{equation}
where $d\Omega ^{2}=d\theta ^{2}+\sin ^{2}\theta d\phi ^{2}$, which simplifies to $d\Omega ^{2}=d\phi ^{2}$
along the galactic plane $\theta=\pi /2$. From the Lagrange equations it follows that we have two constants of
motion, namely, the energy $E$ per unit mass, and the angular momentum $l$ per unit mass, given by $E=e^{\nu (r)}c^3\dot{t}$ and
$l=cr^{2}\dot{\phi}$, respectively. The normalization condition for the four-velocity $u^{\mu }u_{\mu }=1$ gives
$1=e^{\nu \left( r\right) }c^2\dot{t}^{2}-e^{\lambda (r)}\dot{r}^{2}-r^{2}\dot{\phi}^{2}$, from which, with the
use of the constants of motion, we obtain the energy of the particle as
\begin{equation}
\frac{E^{2}}{c^2}=e^{\nu +\lambda }\dot{r}^{2}+e^{\nu }\left( \frac{l^{2}}{c^2r^{2}}%
+1\right) .  \label{energy}
\end{equation}

From Eq.~(\ref{energy}) it follows that the radial motion of the test particles is analogous to that of particles in
Newtonian mechanics, having a  velocity $\dot{r}$, a position dependent effective mass $m_{\rm eff}=2e^{\nu +\lambda }$, and an energy $E^{2}$, respectively. In addition to this, the test particles move in an effective potential provided by the following relationship
\begin{equation}
V_{\rm eff}\left( r\right) =e^{\nu (r)}\left( \frac{l^{2}}{c^2r^{2}}+1\right) .
\end{equation}

The conditions for circular orbits, namely, $\partial V_{\rm eff}/\partial r=0$ and $\dot{r}=0$ lead to
\begin{equation}\label{cons1}
l^{2}=\frac{c^2}{2}\frac{r^{3} \nu ^{\prime } }{1-r\nu ^{\prime }/2},
\end{equation}
and
\begin{equation}\label{cons2}
\frac{E^{2}}{c^4}=\frac{e^{\nu } }{1-r\nu ^{\prime }/2},
\end{equation}
respectively.

Note that the spatial three-dimensional  velocity is given by \cite{LaLi}
\begin{equation}
v^{2}(r)=e^{-\nu }\left[ e^{\lambda }\left( \frac{dr}{dt}\right)
^{2}+r^{2}\left( \frac{d\Omega }{dt}\right) ^{2}\right] .
\end{equation}
For a stable circular orbit $dr/dt=0$, and the tangential velocity of the
test particle can be expressed as
\begin{equation}
v_{tg}^{2}(r)=e^{-\nu }r^{2}\left( \frac{d\Omega }{dt}\right) ^{2}=e^{-\nu }r^{2}\left( \frac{d\Omega }{ds}\right) ^{2}\left(\frac{ds}{dt}^2\right).
\end{equation}
In terms of the conserved quantities, and along the galactic plane $\theta =\pi /2$, the angular velocity is given by
\begin{equation}
\frac{v_{tg}^{2}(r)}{c^2}=c^2\frac{e^{\nu }}{r^{2}}\frac{l^{2}}{E^{2}}\,,
\end{equation}
and taking into account Eqs.~(\ref{cons1}) and (\ref{cons2}), we finally obtain the following relationship \cite{Nuc01}
\begin{equation}
\frac{v_{tg}^{2}(r)}{c^2}=\frac{r \nu ^{\prime }}{2}.  \label{vtg}
\end{equation}

Therefore, once the tangential velocity of test particles is known, the metric function $\nu(r)$ outside the galaxy can
be obtained as
\be
\nu (r)=2\int{\frac{v_{tg}^2(r)}{c^2}\frac{dr}{r}}.
  \label{metricnu}
\ee

The tangential velocity $v_{tg}/c$ of gas clouds moving like test particles around the center of a galaxy is not directly
measurable, but can be inferred from the redshift $z_{\infty }$ observed at spatial infinity, for which
$1+z_{\infty}=\exp \left[ \left( \nu _{\infty }-\nu \right) /2 \right] \left( 1\pm v_{tg}/c\right) /\sqrt{1-v_{tg}^{2}/c^2}$ \cite{Nuc01}.
Due to the non-relativistic velocities of the gas clouds, with $v_{tg}/c\leq \left( 4/3\right) \times 10^{-3}$, we observe
that $v_{tg}/c\approx z_{\infty }$, as the first part of a geometric series. The observations show that at distances large
enough from the galactic center the tangential velocities assume a constant value, i.e., $v_{tg}/c\approx $ constant \cite{dm}. In the latter regions of the constant tangential velocities, Eq. (\ref{metricnu}) can be readily integrated to provide the following metric tensor component
\be\label{nu}
e^{\nu }=\left(\frac{r}{R_{\nu }}\right)^{2v_{tg}^2/c^2}\approx 1+2\frac{v_{tg}^2}{c^2}\ln\left(\frac{r}{R_{\nu }}\right),
\ee
where $R_{\nu }$ is an arbitrary constant of integration. If we match the metric given by Eq.~(\ref{nu}) with the Schwarzschild metric on the surface of the galactic baryonic matter distribution, having a radius $R_B$, $\left.e^{\nu}\right|_{r=R_B}=1-2GM_B/c^2R_B$, we obtain the following relationship
\be
R_{\nu}=\frac{R_B}{\left(1-2GM_B/c^2R_B\right)^{c^2/2v_{tg}^2}}.
\ee

An important physical requirement for the circular orbits of the test particle around galaxies is that they
must be stable. Let $r_{0}$ be the radius of a circular orbit and consider a perturbation of it of the form
$r=r_{0}+\delta $, where $\delta \ll r_{0}$ \citep{La03}. Taking expansions of $V_{\rm eff}\left(
r\right) $ and $\exp \left( \nu +\lambda \right) $  about $r=r_{0} $, it follows from Eq.~(\ref{energy})
that
\begin{equation}
\ddot{\delta}+\frac{1}{2} e^{\nu \left( r_{0}\right)
+\lambda \left( r_{0}\right) }V_{\rm eff}^{\prime \prime }\left( r_{0}\right)
\delta =0.
\end{equation}
The condition for stability of the simple circular orbits requires $V_{\rm eff}^{\prime \prime }\left(
r_{0}\right) >0$ \citep{La03}. Hence, with the use of the condition $V_{\rm eff}^{\prime}\left (r_0\right)=0$, we obtain the condition of the stability of the orbits as $\left[3\nu '+r\nu ''>r\nu '^2/2\right]|_{r=r_0}$. By taking into account Eq.~(\ref{vtg}), it immediately follows that for massive test particles whose velocities are determined by the $g_{00}$ component of the metric tensor only the stability condition of the circular orbits is always satisfied.

%-----------------------------------------------------------------------
\section{Galactic geometry and tangential velocity curves in hybrid metric-Palatini gravity}\label{sect4}
%-----------------------------------------------------------------------

The rotation curves only determine one, namely $\nu (r)$, of the two unknown metric functions, $\nu (r)$ and
$\lambda (r)$, which are required to describe the gravitational field of the galaxy. Hence, in order to determine
$\lambda(r)$ we proceed to solve the gravitational field equations for the hybrid metric-Palatini gravitational theory
 outside the baryonic matter distribution. This allows us to take all the components of the
ordinary matter stress-energy tensor as being zero. Taking into account the stress-energy tensor for the equivalent scalar field representation of hybrid metric-Palatini gravity  the gravitational field equations describing the geometry of the galactic halo take the form
\bea \label{f1b}
-e^{-\lambda }\left( \frac{1}{r^{2}}-\frac{\lambda ^{\prime }}{r}\right) +\frac{1}{r^{2}}&=&\left(\rho _{\phi}+p_{\phi }\right)u_{(\phi ) t}u^t_{(\phi)}-p_{\phi}
    \nonumber\\
&&\hspace{-1.25cm} +q^{(\phi)t}u_{(\phi)t}+q^{(\phi)}_tu_{(\phi)}^t+S^{(\phi)t}_t,
\eea
\bea\label{f2b}
e^{-\lambda }\left( \frac{\nu ^{\prime }}{r}+\frac{1}{r^{2}}\right) -\frac{1
}{r^{2}} &=&-\left(\rho _{\phi}+p_{\phi }\right)u_{(\phi ) r}u^r_{(\phi)}+p_{\phi}
   \nonumber\\
&&\hspace{-1.25cm}  -q^{(\phi)r}u_{(\phi)r}-q^{(\phi)}_ru_{(\phi)}^r-S^{(\phi)r}_r,
\eea
\bea
&&\frac{1}{2}e^{-\lambda }\left( \nu ^{\prime \prime }+\frac{\nu ^{\prime 2}}{%
2}+\frac{\nu ^{\prime }-\lambda ^{\prime }}{r}-\frac{\nu ^{\prime }\lambda
^{\prime }}{2}\right)  =- S^{(\phi)\phi }_{\phi }
\nonumber\\
&&\hspace{-0.65cm} -\left(\rho _{\phi}+p_{\phi }\right)u_{(\phi ) \phi }u^{\phi }_{(\phi)}+p_{\phi}-
q^{(\phi)\phi }u_{(\phi)\phi }-q^{(\phi)}_{\phi }u_{(\phi)}^{\phi }
\,,
\eea
where there is no summation upon the pair of indices $\left(t,r,\phi \right)$.

%-----------------------------------------------------------------------
\subsection{Weak field limit of the gravitational field equations}
%-----------------------------------------------------------------------

The weak field limit of the gravitational theories at the Solar System level is usually obtained by using isotropic coordinates. However, it is useful to apply Schwarzschild coordinates in studying exact solutions and in the context of galactic dynamics. In the following, we will adopt in our analysis the Schwarzschild coordinate system. We assume that the gravitational field inside the halo is weak, so that $\nu (r)\sim \lambda (r)\sim (v_{tg}/c)^2$,
which allows us to linearise the gravitational field equations retaining only terms linear in $(v_{tg}/c)^2$. Moreover,
we assume that the scalar field $\phi $ is also weak, so that $\phi \ll 1$. By representing the scalar field as $\phi
=\phi _0+\varphi $, where $\varphi \ll 1$ is a small perturbation around the background value $\phi _0>0$ of the
field, in the first order of perturbation, the scalar field potential $V(\phi )$ and its derivative with respect to $\phi $
can be represented as
\be
V(\phi )=V\left(\phi _0+\varphi \right)\approx V\left(\phi _0\right)+V'\left(\phi _0\right)\varphi +....,
\ee
and
\be
V'(\phi )\approx V'\left(\phi _0\right)+  V''(\phi_0) \varphi\ ,
\ee
respectively.  In the linear approximation we have $e^{\nu +\lambda }\approx 1$. Therefore the effective Klein-Gordon type equation of the scalar field, Eq.~(\ref{eq:evol-phi}) takes the form
\be\label{yuk}
\left(\nabla^2 -\frac{1}{r_\varphi ^2}\right)\varphi =0,
\ee
where a constant on the right-hand side of this equation has been absorbed into a redefinition of $\phi_0$, and the following parameter has been defined for notational simplicity
\be
\frac{1}{r_{\varphi}^2 }=\frac{1}{3}\left[2V\left(\phi _0\right)-V'\left(\phi _0\right)-\phi _0(1+\phi_0)V''\left(\phi _0\right)\right] \ .
\ee
From a physical point of view $r_{\varphi }$ represents (in natural units) the inverse of the mass $m_{\varphi}$ of the particle associated with the scalar field, $r_{\varphi}=1/m_{\varphi}$. The hybrid metric-Palatini gravity theory can pass the Solar System observational
constraints even if the scalar field is very light, that is, $m_{\varphi}$ is very small \cite{fX}. Within this linear approximation the stress-energy tensor of the scalar field is given by
\be
T_{\mu \nu }^{(\phi)}=\frac{1}{\kappa ^2}\left[\nabla _{\mu }\nabla _{\nu }\varphi +\left(\alpha \varphi +\beta \right)g_{\mu \nu}\right],
\ee
where $\alpha$ and $\beta$ are defined by
\be
\alpha =\frac{1}{r_{\varphi }^2}- \frac{1}{2}V'\left(\phi _0\right)\ , \qquad \beta =-\frac{1}{2}V\left(\phi _0\right).
\ee

Therefore the linearized gravitational field equations take the form
\bea\label{f1}
\frac{1}{r^2}\frac{d}{dr}\left(r\lambda\right)=\alpha \varphi +\beta &=&\rho ^{(\rm eff)},
  \\
%\ee
%\be
\label{f2}
-\frac{\nu '}{r}+\frac{\lambda }{r^2}=\varphi ''+\alpha \varphi +\beta &=&-p_r^{(\rm eff)},
   \\
%\ee
%\be
\label{f3}
-\frac{1}{2}\left(\nu ''+\frac{\nu '-\lambda '}{r}\right)=\alpha \varphi +\beta &=&-p_{\perp}^{(\rm eff)}.
\eea

%-----------------------------------------------------------------------
\subsection{Tangential velocity of test particles in hybrid metric-Palatini gravity}
%-----------------------------------------------------------------------

Using spherical symmetry, Eq.~(\ref{yuk}) takes the form
\be
\frac{1}{r}\frac{d^2}{dr^2}r\varphi -\frac{1}{r_{\phi }^2}\varphi =0 \ ,
\ee
which yields the following general solution
\be
\varphi (r)=\Psi _0\frac{e^{-r/r_{\varphi}}}{r},
\ee
where $\Psi _0$ is an integration constant. Comparing this expression with the results obtained in \cite{fX} for the weak-field limit (taking into account the transformation from isotropic to Schwarzschild coordinates), we find that
\be\label{Psi0}
\Psi_0=-\frac{2GM_B}{c^2}\phi_0\frac{e^{R_B/ r_\varphi }}{3}<0,
\ee
where $M_B$ and $R_B$ are the mass and the radius of the galactic baryonic distribution, respectively.

  Eq.~(\ref{f1}) can be immediately integrated to provide
\bea
\lambda (r)&=&\frac{\lambda_0}{r}+\frac{1}{r}\int ^r{\left(\alpha \varphi +\beta \right)\tilde{r}^2d\tilde{r}}
     \nonumber  \\
&=&\frac{\lambda_0}{r}+\frac{\beta}{3}r^2-\frac{\alpha r_\varphi^2 \Psi_0 e^{-r/r_\varphi}}{r}\left(1+\frac{r}{r_\varphi}\right) \ ,
\eea
where $\lambda_0$ is an integration constant. Comparing again with the results obtained in \cite{fX} for the weak-field limit, we find that $\lambda_0=2GM/c^2$.
The tangential velocity of the test particles in stable circular orbits moving in the galactic halo can be derived immediately from Eq.~(\ref{f2}), and is given by
\be
\frac{v_{tg}^2}{c^2}=\frac{r\nu '}{2}=\frac{\lambda }{2}-r^2\frac{\varphi ''}{2}-\frac{\alpha }{2}r^2\varphi-\frac{\beta}{2}r^2,
\ee
which in terms of the solutions found above becomes
\bea
\frac{v_{tg}^2}{c^2}&=&\frac{V_0}{6} r^2 + \frac{GM_B}{c^2r}-\frac{\Psi_0 e^{-r/r_\varphi}}{2r}\times \nonumber\\
&&\left[\left(1+\frac{r}{r_\varphi}\right)(2+\alpha r_\varphi^2)+\frac{r^2}{r_\varphi^2}(1+\alpha r_\varphi^2)\right],
\eea
where $V_0=-\beta =V\left(\phi _0\right)/2$.
The term proportional to $r^2$ corresponds to the cosmological background, namely the de Sitter geometry, and we assume that it has a negligible contribution on the tangential velocity of the test particles at the galactic level.

On the surface of the baryonic matter distribution the tangential velocity must satisfy the boundary condition
\be
\frac{v_{tg}^2\left(R_B\right)}{c^2}\approx \frac{GM_B}{c^2R_B},
\ee
 which, with the use of Eq.~(\ref{Psi0}), gives the following constraint on the parameters of the model,
\be
\left(1+\frac{R_B}{r_\varphi}\right)(2+\alpha r_\varphi^2)+\frac{R_B^2}{r_\varphi^2}(1+\alpha r_\varphi^2)\approx 0.
\ee
In order to satisfy the above condition would require that $-2<\alpha r_{\varphi }^2<-1$, or, equivalently,
\be
V'\left(\phi _0\right)>0,
\ee
and
\be
2<\frac{1}{2}V'\left(\phi _0\right)r_{\varphi }^2<3,
\ee
respectively.

In the regions near the galactic baryonic matter distribution, where $R_B\leq r \ll r_{\varphi}$, we have $e^{-r/r_{\varphi}}\approx 1$, to a very good approximation. Hence in this region the tangential velocity can be approximated as
\bea
\frac{v_{tg}^2}{c^2}&\approx&\frac{2GM_B-c^2\Psi_0\left(\alpha r_{\varphi}^2+2\right)}{2c^2r}-\Psi_0\frac{\alpha r_{\varphi }^2+2}{2r_{\varphi }}\nonumber\\
&&-\frac{\Psi _0}{2r_{\varphi }^2}\left(1+\alpha r_{\varphi }^2\right)r, \qquad R_B\leq r \ll r_{\varphi}.
\eea

If the parameters of the model satisfy the condition
\bea
2GM_B-c^2\Psi_0\left(\alpha r_{\varphi}^2+2\right)\approx 0,
\eea
the term proportional to $1/r$ becomes negligible, while for small values of $\Psi _0$, and $\left|\alpha r_{\varphi ^2}\right|\approx 1$, the term
proportional to $r$ can also be neglected. Therefore for the tangential velocity of test particles rotating in the
galactic halo we obtain
\be
\frac{v_{tg}^2}{c^2}\approx -\Psi_0\frac{\alpha r_{\varphi }^2+2}{2r_{\varphi}}\approx -\frac{\Psi _0\alpha
r_{\varphi}}{2},  \qquad R_B\leq r \ll r_{\varphi}.
\ee
Since according to our assumptions, $r_{\varphi} \gg 1$, the coefficient $\alpha $ can be approximated as $\alpha
\approx -V'\left(\phi _0\right)/2$, which provides for the rotation curve, in the constant velocity region, the
following expression
\be
\frac{v_{tg}^2}{c^2}\approx \frac{\Psi _0V'\left(\phi _0\right) r_{\varphi}}{4}, \qquad R_B\leq r  \ll r_{\varphi}.
\ee
Since $\Psi_0<0$, the scalar field potential must satisfy the condition $V'\left(\phi _0\right)<0$. In the first order of approximation, with $\exp\left(-r/r_{\varphi}\right)\approx1-r/r_{\varphi}$, for the tangential velocity we obtain the expression
\be
\frac{v_{tg}^2}{c^2}\approx \frac{2GM_B-c^2\Psi _0\left(\alpha r_{\varphi}^2+2\right)}{2c^2r}+\frac{\Psi _0}{2r_{\varphi }^2}r+\frac{\Psi_0\left(\alpha r_{\varphi }^2+1\right)}{2r_{\varphi }^2}r^2.
\ee

Alternatively, in general we can write the tangential velocity  as follows,
\bea\label{vfin}
\frac{v_{tg}^2}{c^2}&=&\frac{V_0}{6} r^2 + \frac{GM_B}{c^2r}\Bigg \{1+\frac{2\phi_0}{3}e^{\frac{GM_B/c^2-r}{r_\varphi}}\Bigg [\left(1+\frac{r}{r_\varphi}\right)\times \nonumber\\
&&(2+\alpha r_\varphi^2)+\frac{r^2}{r_\varphi^2}(1+\alpha r_\varphi^2)\Bigg ]\Bigg\}.
\eea
As compared to our previous results, in this representation we have $e^{\frac{GM_B/c^2-r}{r_\varphi}}$ instead of $e^{\frac{R_B-r}{r_\varphi}}$. Since we are working in a regime in which $R_B\ll r_\varphi$, the choice of the constants $R_B$ or $M_B$ does not seem very  relevant, since it just amounts to a rescaling of $\phi_0$. From now on we will also assume that $e^{GM_B/c^2r_\varphi}\approx 1$.

From the above equation we want to find the constraints on the model parameters that arise from the expected behavior at different scales. For that purpose, it is convenient to write the equation, equivalently, as follows:
\bea
\frac{v_{tg}^2}{c^2}&=& \frac{GM_B}{c^2r}\left[1+\frac{2\phi_0}{3}(2+\alpha r_\varphi^2)e^{-\frac{r}{r_\varphi}}\right] + \nonumber\\
&& \hspace{-1.8cm} \frac{GM_B}{c^2r_\varphi} \left(2+\alpha r_\varphi^2\right) e^{-\frac{r}{r_\varphi}}+\frac{GM_B}{c^2r_\varphi} (1+\alpha r_\varphi^2) \left(\frac{r}{r_\varphi}\right) e^{-\frac{r}{r_\varphi}} .
\eea

At intermediate scales, the asymptotic tangential velocity tends to a constant.
If we expand the exponential as $e^{-\frac{r}{r_\varphi}}\approx 1-r/r_\varphi$, then we obtain the following three constraints on the free parameters of the model,
\bea
&&a) \;\;1+\frac{2\phi_0}{3}\left(2+\alpha r_\varphi^2 \right)\approx 0, \\
&&b) \;\;  \left(2+\alpha r_\varphi^2 \right)\left(1-\frac{2\phi_0}{3}\right)\approx C= {\rm constant},\\
&&c) \; \;\frac{GM_B}{c^2r_\varphi}\left(\frac{r}{r_\varphi}\right) \ll |C|.
\eea

With increasing $r$, and by assuming that the condition $r \ll r_{\varphi}$ still holds, the rotation curves will decay, at very large distances from the galactic center, to the zero value.

%-----------------------------------------------------------------------
\section{Astrophysical tests of hybrid metric-Palatini gravity at the galactic level}\label{sect5}
%-----------------------------------------------------------------------

In the present Section, we will present some observational possibilities of directly checking the validity of the hybrid metric-Palatini gravitational model. More specifically, we will first compare the theoretical predictions of the model with a sample of rotation curves of low surface brightness  galaxies, respectively. Then we consider the possibility of observationally determining the functional form of the scalar field $\varphi $ by using the velocity dispersion of stars in galaxies, and the red and blue shifts of gas clouds moving in the galactic halo.

\subsection{Low surface brightness galaxy rotation curves in the hybrid metric-Palatini gravity}

In order to test our results we compare the predictions of our model with the observational data on the galactic
rotation curves, obtained for a sample of low surface luminosity galaxies in \cite{LSB1}. Generally, in a realistic situation, a galaxy consists of a distribution of baryonic (normal) matter, consisting of stars
of mass $M_{star}$, ionized gas of mass $M_{gas}$, neutral hydrogen of mass $M_{HI}$ etc., and the ``dark matter'' of mass $M_{DM}$, which, in the present model, is generated by the extra contributions to the total energy-momentum, due to the contribution of the effective scalar field. Hence the total mass of the galactic baryonic matter is $M_B= M_{star}+ M_{gas}+ M_{HI}+...$. As for the distribution of the baryonic mass, we assume that it is concentrated into an inner core of radius $r_c$, and that its mass profile $m_B(r)$ can be described by the simple relation
\be
m_B (r) = M_B\left(\frac{r}{r+r_c}\right)^{3\beta },  \qquad  r\leq R_B, R_B \gg r_c,
\ee
where $\beta = 1$ for High Surface Brightness galaxies (HSB) and $\beta = 2$ for Low Surface Brightness (LSB) and dwarf galaxies,
respectively \cite{mprof}. For $r=R_B$ we have $m\left(R_ B\right)\approx M_B$.  By representing the coefficient $\alpha $ as $\alpha =\alpha _0/r_{\varphi }^2$, where $\alpha _0=1-r_{\varphi }^2V'\left(\phi _0\right)/2>>1$, from Eq.~(\ref{vfin}) we obtain the tangential velocity of the massive particles in stable galactic circular orbits as
\be
v_{tg}^2=v_{Kepl}^2+\frac{2\alpha _0\phi _0}{3}v_{Kepl}^2e^{-r/r_{\varphi}}\left(1+\frac{r}{r_{\varphi}}+\frac{r^2}{r_{\varphi }^2}\right),r\geq R_B,
\ee
where $v_{Kepl}^2=GM_B/r$. Hence, by also taking into account the baryonic matter contribution, we obtain the total tangential velocity of a massive test particle  as
\bea\label{comp}
v_{tgtot}^2\left({\rm km/s}\right)&=&4.33\times 10^4\, \frac{M_B\left(10^{10}M_{\odot}\right)}{r\;({\rm kpc})}
  \left(\frac{r}{r+r_c}\right)^{6 }
 \nonumber\\
&&+2.77\times 10^4\, \alpha _0 \, \phi _0\, \frac{M_B\left(10^{10}M_{\odot}\right)}{r\;({\rm kpc})} 
  \nonumber  \\
 && 
\times \left(1+\frac{r}{r_{\varphi}}+\frac{r^2}{r_{\varphi }^2}\right), \qquad  r\geq R_B ,
\eea
where we have used the value $\beta =2$.

\begin{figure*}
   \centering
  \includegraphics[width=8cm]{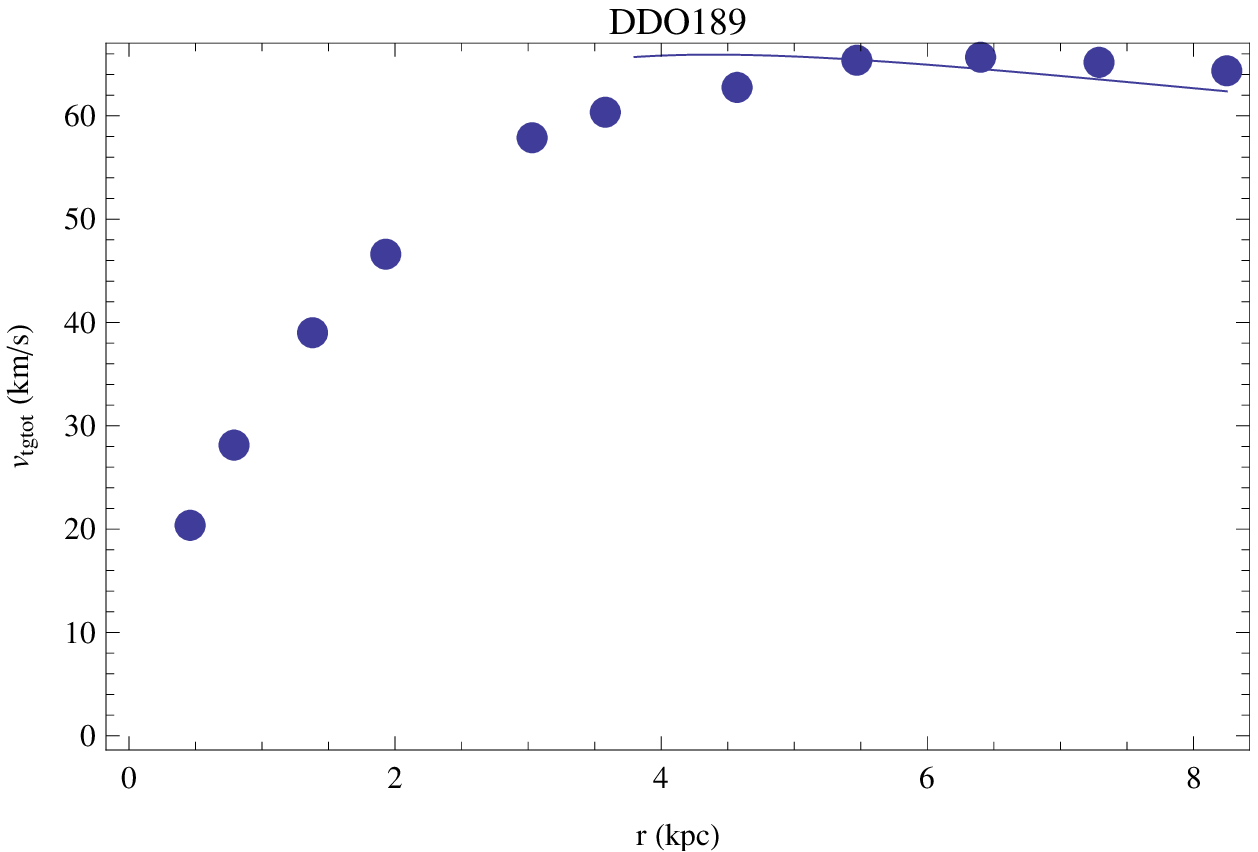}
   \includegraphics[width=8cm]{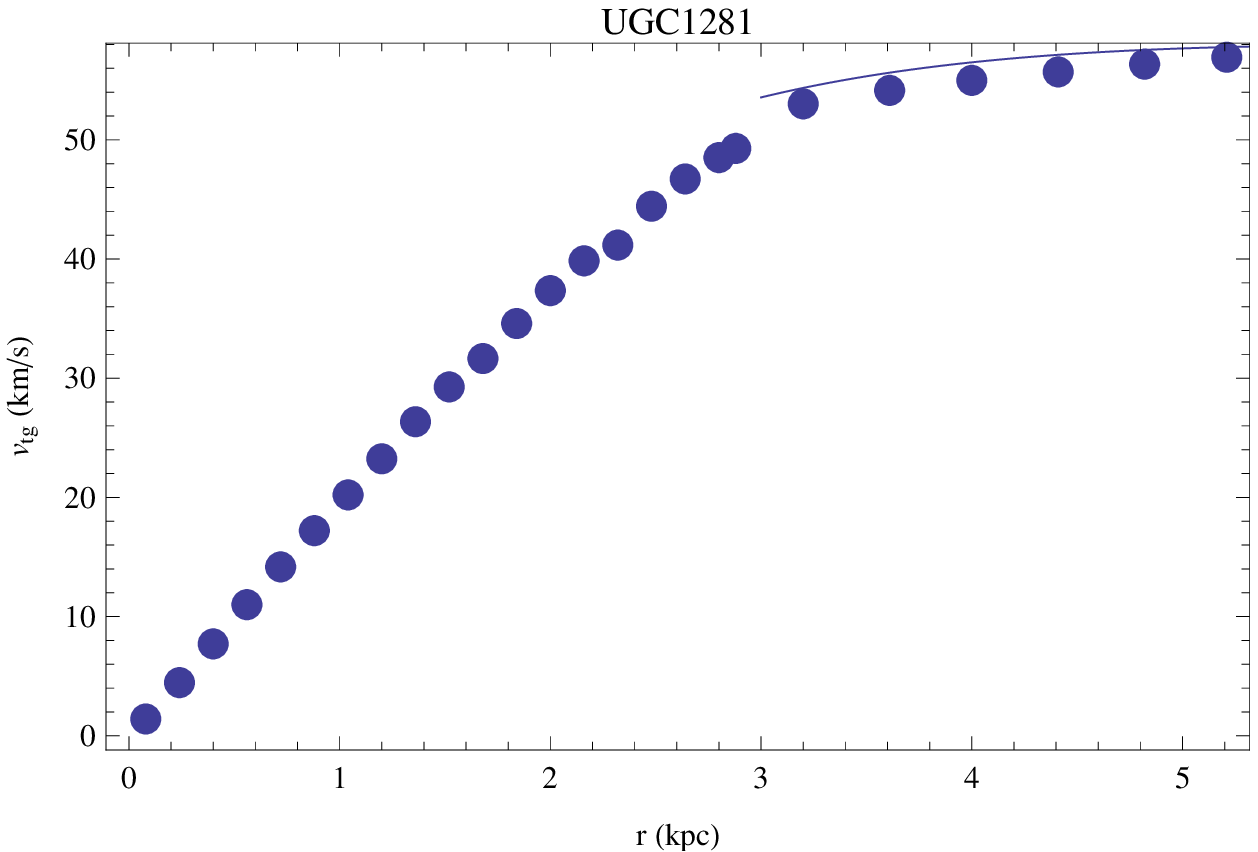}
    \includegraphics[width=8cm]{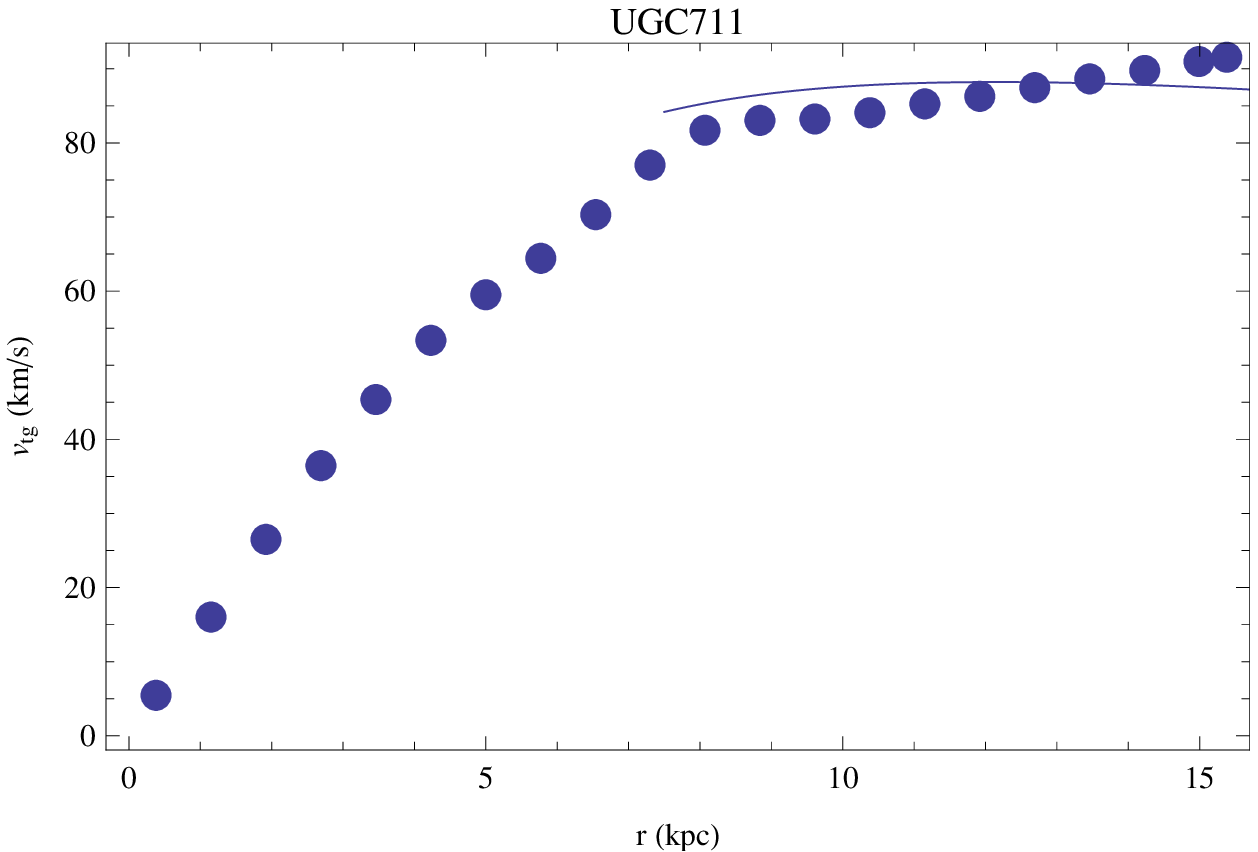}
   \includegraphics[width=8cm]{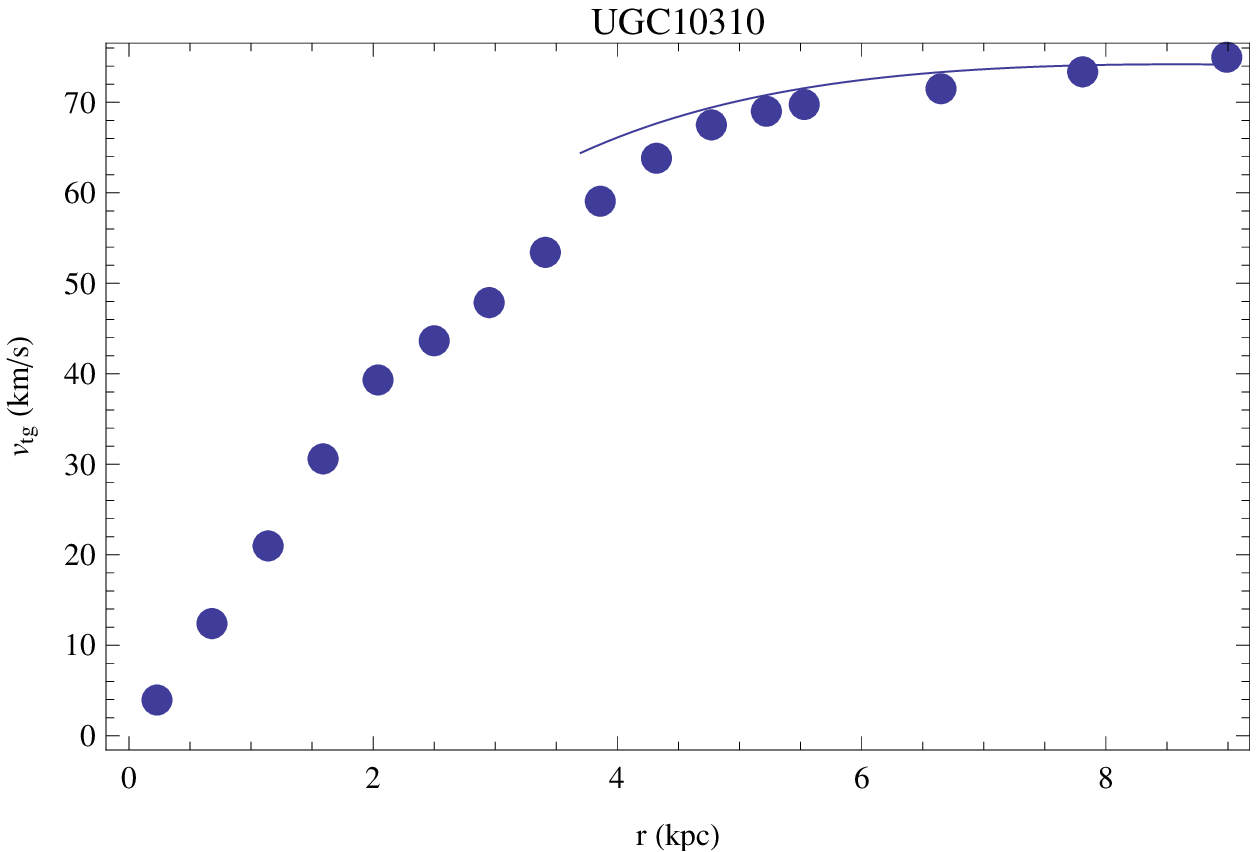}
   \caption{ Comparison of the theoretical predictions of the behavior of the galactic rotation curves in  the constant velocity region (solid curves), and the observational results (points) obtained in \cite{LSB1} for four LSB galaxies. For the universal parameters of the hybrid metric-Palatini gravity model we have adopted the values $\alpha _0=50$, $\phi _0=6\times 10^{-4}$, and $r_{\varphi }=20\;{\rm kpc}$, respectively. The total baryonic mass and the core radius of the  considered galaxies are given by: $M_B=1.05\times 10^{10}M_{\odot}$, $r_c=0.87$ kpc (DDO189), $M_B=1.0\times 10^{10}M_{\odot}$, $r_c=1.16$ kpc (UGC1281), $M_B=5.7\times 10^{10}M_{\odot}$, $r_c=2.4$ kpc (UGC711), and $M_B=2.7\times 10^{10}M_{\odot}$, $r_c=1.69$ kpc (UGC10310), respectively.}
         \label{fig1}
   \end{figure*}

In order to compare the prediction of Eq.~(\ref{comp}) with the observed rotation  curves of the LSB galaxies in the following we assume $V'\left(\phi _0\right)<0$, and we fix the numerical values of the universal parameters $\left(\alpha _0, \phi _0,r_{\varphi}\right)$ as $\left(\alpha _0=50, \phi _0=6\times 10^{-4}, r_{\varphi }=20\;{\rm kpc}\right)$. Then any variability in the behavior of the rotation curves is due to the variation of the baryonic mass of the galaxy $M_B$, and of its baryonic mass distribution in the core, described by $r_c$. In Fig.~\ref{fig1}, we have compared the predictions of Eq.~(\ref{comp}) for the behavior of the rotation curves in the ``dark matter'' region for four LSB galaxies, DDO189, UGC1281, UGC711, and UGC10310, respectively \cite{LSB1}.

The comparison of the predictions of the theoretical model with the observational results show that the contribution of the scalar field energy density to the tangential velocity of the test particles can explain the existence of a constant rotational velocity region around the baryonic matter, without requiring the presence of the dark matter. Of course, in order to gain a better understanding of the behavior of the galactic rotation curves in the hybrid metric-Palatini gravity model, the qualitative approach considered in the present Section must be reconsidered by taking into account more realistic galactic baryonic matter distributions, and a much larger sample of galaxies having different morphologies.
%-----------------------------------------------------------------------
\subsection{Constraining hybrid metric-Palatini gravity with velocity dispersions}
%-----------------------------------------------------------------------

In hybrid metric-Palatini gravity one can formally associate an approximate ``dark matter'' mass profile $M_{DM}(r)$ to the tangential velocity profile, which taking into account Eq. (\ref{f1}), is given by
\begin{equation}\label{darkmass}
\frac{2GM_{DM}(r)}{c^2}=\int_{R_B}^r{\left(\alpha \varphi +\beta \right)r^2dr},
\end{equation}
so that the metric tensor component $\lambda(r)$ can be written as
\be
\lambda(r)=\frac{2GM_{DM}(r)}{r}.
\ee
The effective ``dark matter'' density profile $\rho _{DM}$ is obtained as
\be
 \rho _{DM}(r)=\frac{1}{4\pi r^2}\frac{dM_{DM}(r)}{dr}=\frac{c^2}{2G}\left(\alpha \varphi+\beta \right).
 \ee

In order to observationally constrain $M_{DM}$ and $\rho _{DM}$, we assume
that each galaxy consists of a single, pressure-supported stellar population
that is in dynamic equilibrium and traces an underlying gravitational
potential, which is created due to the presence of the scalar field $\varphi $. By assuming spherical symmetry, the equivalent mass profile
induced by the scalar field (the mass profile of the ``effective dark matter'' halo)
can be obtained from the moments of the stellar distribution function via the Jeans
equation \cite{BT08}
\begin{equation}
\frac{d}{dr}\left[ \rho _{s}\left\langle v_{r}^{2}\right\rangle \right] +
\frac{2\rho _{s}\left( r\right) \beta _{an}(r)}{r}=-\frac{G\rho _{s}M_{DM}(r)}{r^{2}},
\end{equation}%
where $\rho _{s}(r)$, $\left\langle v_{r}^{2}\right\rangle $, and $\beta _{an}
(r)=1-\left\langle v_{\theta }^{2}\right\rangle /\left\langle
v_{r}^{2}\right\rangle $ describe the three-dimensional density, the radial
velocity dispersion, and the orbital anisotropy of the stellar component, where $\left\langle v_{\theta }^{2}\right\rangle$ is the tangential velocity dispersion. With the assumption of constant  anisotropy, $\beta _{an}={\rm constant}$, the Jeans equation can be solved to give $\rho _s$ as  \cite{MaLo05}
\begin{equation}
\rho _{s}\left\langle v_{r}^{2}\right\rangle  =Gr^{-2\beta _{an}
}\int_{r}^{\infty }s^{2(1-\beta _{an} )}\rho _{s}\left( s\right) M_{DM}\left(
s\right) ds.
\end{equation}
With the use of Eq.~(\ref{darkmass}) we obtain for the stellar velocity dispersion the equation
\bea\label{integral}
\rho _{s}\left\langle v_{r}^{2}\right\rangle &\approx & \frac{c^2}{4G}r^{-2\beta _{an}}\int_{r}^{\infty }s^{2(2-\beta _{an})}\rho
_{s}\left( s\right) \times \nonumber\\
&& \times \left\{\int_{R_B}^s{\left[\alpha \varphi (\xi) +\beta\right]\xi^2 d\xi}\right\}ds.
\eea

The ``effective dark matter'' mass profile can be related through the projection along the line of sight  to two observable quantities, the projected stellar density $I(R)$, and to the stellar velocity dispersion $\sigma _p(R)$, respectively,  according to the relation \cite{BT08}
\begin{equation}
\sigma _{P}^{2}(R)=\frac{2}{I(R)}\int_{R}^{\infty }\left( 1-\beta _{an}\frac{R^{2}
}{r^{2}}\right) \frac{\rho _{s}\left\langle v_{r}^{2}\right\rangle r}{\sqrt{
r^{2}-R^{2}}}dr.
\end{equation}

Given a projected stellar density model $I(R)$, one recovers the
three-dimensional stellar density from \cite{BT08}
\begin{equation}
\rho _{s}(r)=-  \frac{1}{\pi} \int_{r}^{\infty } \,\frac{dI}{dR}  \left( R^{2}-r^{2}\right) ^{-1/2}dR \,.
\end{equation}
Therefore, once the stellar density profile $I(R)$, the stellar velocity dispersion $\left\langle v_{r}^{2}\right\rangle$,  and the quantities $\alpha $, $\beta $, $R_B$ and $M_B$, determining the geometry of the space-time outside the baryonic matter distribution,
 are known, with the use of the integral equation Eq.~(\ref{integral}) one can constrain  the explicit functional form of the scalar field $\varphi$,  the two free parameters of the model, $\Psi _0$ and $r_{\varphi}$, as well as  the equivalent mass and density profiles induced by the presence of the scalar field.
  The simplest analytic projected density profile is the Plummer profile \cite{BT08}, given by $I(R)={\cal L}\left(\pi r_{half}^2\right)^{-1}\left(1+R^2/r_{half}^2\right)^{-2}$, where ${\cal L}$ is the total luminosity, and $r_{half}$ is the projected half-light radius (the radius of the cylinder that encloses half of the total luminosity).

%---------------------------------------------------------------
\subsection{Red and blue shifts of the electromagnetic radiation}
%---------------------------------------------------------------

The rotation curves of spiral galaxies are inferred from the astrophysical observations of the red and
blue shifts of the radiation emitted by gas clouds moving in circular orbits on both sides of the central
region in the galactic plane. The light signal travels on null geodesics in the galactic geometry with
tangent $k^{\mu }$. We may, without a significant loss of generality, restrict $k^{\mu }$ to lie in the
equatorial plane $\theta =\pi /2$, and evaluate the frequency shift for a light signal emitted from the
observer $O_{E}$ in circular orbit in the galactic halo, and detected by the observer $O_{\infty }$
situated at infinity. The frequency shift associated to the emission and detection of the light signal from
the gas cloud is defined as
\begin{equation}
z=1-\frac{\omega _{E}}{\omega _{\infty }},
\end{equation}
where $\omega _{I}=-k_{\mu }u_{I}^{\mu }$, and the index $I$ refers to emission ($I=E$) or detection
($I=\infty $) at the corresponding space-time point \cite{Nuc01,La03}. We can associate with light
propagation two frequency shifts, corresponding to maximum and minimum values,   in the same and
opposite direction of motion of the emitter, respectively. From an astrophysical point of view such shifts
are frequency shifts of a receding or approaching gas cloud, respectively. In terms of the tetrads
$e_{(0)}=e^{-\nu /2}\partial /\partial t$, $e_{(1)}=e^{-\lambda /2}\partial /\partial r$,
$e_{(2)}=r^{-1}\partial /\partial \theta $, $e_{(3)}=\left( r\sin \theta \right) ^{-1}\partial
/\partial \phi $, the frequency shifts can be represented as \cite{Nuc01}
\begin{equation}\label{60}
z_{\pm }=1-e^{\left[ \nu _{\infty }-\nu \left( r\right) \right] /2}\left(
1\mp v\right) \Gamma ,
\end{equation}
where $v=\left[ \sum_{i=1}^{3}\left( u_{(i)}/u_{(0)}\right) ^{2}\right]^{1/2}$, with $u_{(i)}$ the
components of the particle's four velocity along the tetrad (i.e., the velocity measured by an Eulerian
observer whose world line is tangent to the static Killing field). In Eq.~(\ref{60}), $\Gamma =\left( 1-
v^{2}\right)^{-1/2}$ is the usual Lorentz factor, and $\exp \left( \nu _{\infty }\right)$ represents the
value of $\exp \left[ \nu \left( \left( r\right) \right) \right] $ for $r\rightarrow \infty $. In the case of
circular orbits in the $\theta =\pi /2$ plane, we obtain
\begin{equation}
z_{\pm }=1-e^{\left[ \nu _{\infty }-\nu \left( r\right) \right] /2}\frac{%
1\mp \sqrt{r \nu ^{\prime } /2 }}{\sqrt{1-r \nu ^{\prime } /2 }}.
\end{equation}

It is convenient to define two other quantities, $z_{D}=\left(z_{+}-z_{-}\right) /2$, giving the differences in the Doppler shifts for the receding and approaching gas clouds, and $z_{A}=\left(
z_{+}+z_{-}\right) /2$, representing the mean value of the Doppler shifts, respectively \cite{Nuc01}. These redshift factors are given by
\begin{equation}\label{zd}
z_{D}\left( r\right) =e^{\left[ \nu _{\infty }-\nu \left( r\right) \right]
/2}\frac{\sqrt{r\nu ^{\prime } /2
 }}{\sqrt{1-r \nu ^{\prime }/2 }},
\end{equation}
and
\begin{equation}\label{za}
z_{A}\left( r\right) =1-\frac{e^{\left[ \nu _{\infty }-\nu \left( r\right) %
\right] /2}}{\sqrt{1-r \nu ^{\prime } /2 }},
\end{equation}
respectively, and they  can be easily connected to the astrophysical observations \citep{Nuc01}. $z_{A}$
and $z_{D}$ satisfy the relation $\left( z_{A}-1\right)^{2}-z_{D}^{2}=\exp \left[ 2\left( \nu _{\infty }-\nu
\left( r\right)\right) \right] $, and thus in principle, the metric tensor component $\exp \left[ \nu  \left(
r\right)  \right] $ can be directly  determined from observations. From Eq.~(\ref{f2}) we obtain
\be
r\nu '=\frac{2GM_{DM}}{c^2r}-\left(\varphi ''+\alpha \varphi +\beta\right)r^2,
\ee
and
\be
\nu (r)=\int_{R_B}^r{\left[\frac{2GM_{DM}}{c^2r^2}-\left(\varphi ''+\alpha \varphi
+\beta)\right)r\right]dr},
\ee
respectively. By substituting these expressions of the metric tensor and of its derivative in
Eqs.~(\ref{zd}) and (\ref{za}), in principle we obtain a direct observational test of the galactic geometry,
of the functional form of the scalar field, and, implicitly, of the hybrid metric-Palatini gravitational
model.

%-----------------------------------------------------------------------
\section{Discussions and final remarks}\label{sect6}
%-----------------------------------------------------------------------

The behavior of the galactic rotation curves, especially their constancy, and the mass deficit in clusters
of galaxies, continues to pose a major challenge to present day physics. It is essential to have a better
understanding of some of the intriguing phenomena associated with them, such as their universality,
the very good correlation between the amount of dark matter and the luminous matter in the galaxy, as
well as the nature of the dark matter particle, if it really does exist. To explain these intriguing
observations, the commonly adopted models are based on exotic, beyond the standard model, particle
physics in the framework of Newtonian gravity, or of some extensions of general relativity.

In the present paper, we have considered the observational implications of the model proposed in
\cite{fX}, and proposed an alternative view to the dark matter problem, namely, the possibility that the
galactic rotation curves and the mass discrepancy in galaxies can naturally be explained in gravitational
models in which  an $f(\R)$ term, constructed within the framework of the Palatini formalism, is added
to the metric Einstein-Hilbert Lagrangian. The extra-terms in the gravitational field equations, which
can be described as a function of an equivalent scalar field, modify, through the metric tensor
components, the equations of motion of test particles, and induce a supplementary gravitational
interaction, which can account for the observed behavior of the galactic rotation curves. Due to the
presence of the scalar field, the rotation curves show a constant velocity region, which decay to zero at
large distances from the galactic center, a behavior which is perfectly consistent with the observational
data \cite{dm}, and is usually attributed to the existence of dark matter. By using the weak field
limit of the gravitational field equations,  the rotation curves  can be completely reconstructed as
functions of the scalar field, without any supplementary assumption. If the galactic rotation velocity
profiles are known from observations,  the galactic metric can be derived theoretically, and the scalar
field function can be reconstructed exactly over the entire mass distribution of the galactic halo.

The formalism developed in the present paper could also be extended to the case of the galaxy clusters. The latter are cosmological structures consisting of hundreds or thousands of galaxies. We emphasize that the analysis of the geometric properties of the galaxy clusters can also be done in weak field approximation considered in the present paper. The comparison of the observed velocity dispersion profiles of the galaxy clusters and the velocity dispersion profiles predicted by the hybrid metric-Palatini gravity model can provide a powerful method for the observational test of the theory, and for observationally discriminating between the different modified gravity theoretical models.

The nature and dynamics of the cosmological evolution can be investigated by using a variety of cosmological observations. One of the important methods for the study of the cosmic history relies on extracting the baryon acoustic oscillations (BAO) in the high-$z$ galaxy power spectrum. The baryon acoustic oscillations  imprint a characteristic scale on the galaxy distribution that acts as a standard ruler. The origin of the BAO in the matter
power spectrum can be understood as the velocity fluctuation of the baryonic fluid at the decoupling time.
The characteristic scale of the baryon oscillation is determined by the sound speed and horizon at decoupling, which is a function of the total matter and baryon densities \cite{BAO}. Since this scale can be measured in both the transverse and radial directions, the BAO yields both the angular diameter distance, and the Hubble parameter at that redshift. Therefore, the precise measurement of the BAO scale from the galaxy power spectrum can impose important constraints on the cosmic expansion history. Different expansion histories in modified gravity models shifts the peak positions of oscillations relative to the $\Lambda $CDM model \cite{BAO1}. Therefore the predicted shifts in the BAO can potentially be used to distinguish between the $\Lambda$CDM models and modified gravity models. Thus, by using the BAO analysis it can be shown that  the original Dvali-Gabadadze-Porrati model \cite{DGP} is disfavored by observations, unless the matter density parameter exceeds 0.3 \cite{BAO1}.

The recently released Planck satellite data \cite{Planck}, as combined with the BAO measurements \cite{BAOO} provide strong constraints on the modified $f(R)$ gravity model \cite{Up}. In the $f(R)$ modified gravity models the lensing amplitude return to be compatible with $A_L = 1$ at 68\% confidence limit (c.l.) if one consider the Planck or Planck combined with the Hubble Space Telescope measurement data, and even at 95\% c.l. if we consider Planck data combined with BAO data. Moreover, in the framework of the considered $f(R)$ models the standard value of the lensing amplitude $A_L = 1$ is in agreement with the Planck measurements, oppositely to what happens in the $\Lambda $CDM scenario.

The study of the BAO and of the CMB data can provide very powerful and high precision constraints in discriminating between the hybrid metric-Palatini gravity model, and alternative gravity models, as well as the standard dark matter model. Weak gravitational lensing, whereby galaxy images are altered due to the gravitational field influence of the mass along the line of sight, is a powerful probe of the dark matter in cosmology, with promising results obtained in recent years \cite{Lens}. In standard general relativity, the weak lensing distortion field provides a direct tracer of the underlying matter distribution. However, modifications to gravity theory can conceivably alter the way that mass curves spacetime, and thus the way that null geodesics behave in a given matter distribution. The possibilities of constraining modified gravity theories with weak lensing were considered recently in \cite{modlens}. Due to its sensitivity to the growth rate of the structure, weak lensing can be very useful to constrain modified gravity theories, and to distinguish between various modified gravity  and standard dark matter models,  when combined with CMB observations. Future weak lensing surveys as Euclid can constrain modified gravity models, as those predicted by scalar-tensor and $f(R)$ theories. Since the hybrid metric-Palatini gravity model can be formulated in terms of an equivalent scalar-tensor theory, the analysis of the future weak lensing observational data, such as those provided by the Euclid mission, may provide a powerful method to observationally constrain the free parameters of this theoretical model. Even that the effective scalar field of the hybrid metric-Palatini gravity model provides a gravitational ``mass'' equivalent to the dark matter, due to the specific functional form and numerical values of the model parameters, its imprint on the weak lensing properties on a cosmological scale is different from that of the standard dark matter.

In the present model all the relevant physical quantities, including the ``dark mass'' associated to the equivalent scalar-tensor description,  and which plays the role of dark matter, its corresponding density profile, as well as the scalar field and its potential, are expressed in terms of observable parameters -- the tangential velocity, the baryonic (luminous) mass, the Doppler frequency shifts of test particles moving around the galaxy, and the velocity dispersions of the stars. Therefore, this opens the possibility of directly testing the modified gravitational models with Palatini type $f(\R)$ terms added to the gravitational action by using direct astronomical and astrophysical observations at the galactic or extra-galactic scale. In this paper we have provided some basic theoretical ideas, which, together with the virial theorem considered in \cite{Capozziello:2012qt}, are the necessary tools for the in depth comparison of the predictions of the hybrid metric-Palatini gravity model with the observational results.

\acknowledgments 
We would like to thank to the anonymous referee for comments and suggestions that helped us to significantly improve our manuscript. SC is supported by INFN (iniziative specifiche QGSKY and TEONGRAV). TSK is supported by the Research Council of Norway.  FSNL acknowledges financial support of the Funda\c{c}\~{a}o para a Ci\^{e}ncia e Tecnologia through the grants CERN/FP/123615/2011 and CERN/FP/123618/2011.  GJO is supported by the Spanish grant FIS2011-29813-C02-02, the Consolider Programme CPAN (CSD2007-00042), and the JAE-doc program of the Spanish Research Council (CSIC).

\appendix

\section{Isotropic  versus Schwarzschild coordinates}\label{app}

By convention, the weak-field limit of theories of gravity in the Solar System is discussed using isotropic coordinates. However, exact solutions and galactic dynamics are usually considered in terms of the Schwarzschild coordinates. Since we have already obtained the form of the weak-field limit in \cite{fX}, in order to translate those results to the current problem we just need to transform our metric from isotropic to Schwarzschild coordinates. This will give coherence to these series of papers. What we need to do is just to compare the Schwarzschild-like line element used in the study of the galactic dynamics with the isotropic results, and find the change of coordinates.\\

In Schwarzschild coordinates the linearized line element is
\bea
ds^2&=&e^\nu dt^2-e^\lambda dr^2-r^2d\Omega^2 \approx \nonumber\\
&&(1+\nu)dt^2-(1+\lambda)dr^2-r^2d\Omega^2  .
\eea

In isotropic coordinates the linearized line element is
\begin{equation}
ds^2\approx(\eta_{\mu\nu}+h_{\mu\nu})dx^\mu dx^\nu= (1+h_{00})dt^2-(\delta_{ij}-h_{ij})dx^i dx^j .
\end{equation}
Defining $h_{ij}=\Lambda \delta_{ij}$, it follows that
\begin{equation}
ds^2= (1+h_{00})dt^2-(1-\Lambda)d\rho^2-(1-\Lambda)\rho^2 d\Omega^2\ .
\end{equation}

The comparison of the coordinates gives the relations
\begin{eqnarray}
(1+\lambda)dr^2&=&(1-\Lambda)d\rho^2, \label{81}\\
r^2 &=& (1-\Lambda)\rho^2.
\end{eqnarray}
The second of these equations allows to express $h_{00}(\rho)$ and $\Lambda(\rho)$ in terms of the Schwarzschild coordinate $r$. It also allows us to find an expression for $dr/d\rho$ as follows,
\begin{equation}
\left(\frac{dr}{d\rho}\right)^2=\left(\frac{\rho}{r}\right)^2\left[(1-\Lambda)-\frac{\rho}{2}\frac{d\Lambda}{d\rho}\right]^2.
\end{equation}
Inserting this result in Eq.~(\ref{81}), and expanding to leading order, we find
\begin{equation}
(1+\lambda)=\left[1+{\rho}\frac{d\Lambda}{d\rho}\right].
\end{equation}
Therefore, $\nu=h_{00}$ and $\lambda={\rho}d\Lambda/d\rho $. Taking into account the explicit form of $h_{00}$ and $\Lambda$ \cite{fX},
\begin{eqnarray}
h_{00}^{(2)}(\rho)&=&- \frac{2G_{eff} M}{\rho} -\frac{V_0}{(1+\phi_0)}\frac{\rho^2}{6},\\
\Lambda(\rho)&=& -\frac{2\gamma G_{eff} M}{\rho} +\frac{V_0}{(1+\phi_0)}\frac{\rho^2}{6}  ,
\end{eqnarray}
with
\begin{eqnarray}
G_{eff}&\equiv & \frac{G}{(1+\phi_0)}\left(1-\frac{\phi_0}{3}e^{-m_\varphi \rho}\right), \\
\gamma &\equiv & \frac{\left(1+\frac{\phi_0}{3}e^{-m_\varphi \rho}\right)}{\left(1-\frac{\phi_0}{3}e^{-m_\varphi \rho}\right)}  \ ,
\end{eqnarray}
we find that
\begin{eqnarray}
r^2&=&\rho^2\Bigg[1+\frac{G}{(1+\phi_0)}\frac{2M}{\rho}\left(1+\frac{\phi_0}{3}e^{-m_\varphi \rho}\right)-\nonumber\\
&&\frac{V_0}{\left(1+\phi_0\right)}\frac{\rho^2}{6}\Bigg ].
\end{eqnarray}
Here we can consider some simplifications. As a first approximation we can assume $V_0$ to be very small, which is equivalent to consider scales much smaller than the cosmological horizon. We can also assume $\phi_0$ small, but we keep $m_\varphi \rho$ non-negligible. In this case, we find the same relation between coordinates as in GR, namely,
\be
r^2\approx \rho^2\left(1+\frac{2GM}{\rho}\right),
\ee
 which leads to
 \be
 r\approx  \rho\left(1+\frac{GM}{\rho}\right)=GM+\rho.
 \ee
  With this result, we find
\begin{eqnarray}
&&\nu(r)=\left.h_{00}(\rho)\right|_{\rho=r-GM}, \\
&&\lambda(r) = \left.\frac{G}{(1+\phi_0)}\frac{2M}{\rho}\left[1+\phi_0(1+\rho m_\varphi)\frac{e^{-m_\varphi \rho}}{3}\right]\right|_{\rho=r-GM}.\nonumber\\
\end{eqnarray}

Note that even though at galactic scales we have $r\gg GM$ and $\rho\approx r$, the exponential corrections $\phi_0 e^{-m_\varphi \rho}$ experience a magnification of the {\it apparent} value of $\phi_0$ by an exponential factor:  $(\phi_0 e^{m_\varphi GM})e^{-m_\varphi r}$, i.e., $\phi_0 \to \phi_0 e^{GM/r_\varphi}$.

\end{document}